\documentclass[
prc,%
10pt,%
final,%
notitlepage,%
oneside,%
twocolumn,%
nobibnotes,%
nofootinbib,
superscriptaddress,%
floatfix,%
floatfix,%
showkeys,%
showpacs]%
{revtex4}
\usepackage{color}
\usepackage{amsfonts}
\usepackage{amsbsy}
\usepackage{mathrsfs}
\usepackage{graphicx}
\def\lsim{\mathrel{\rlap{
\lower4pt\hbox{\hskip-3pt$\sim$}}
    \raise1pt\hbox{$<$}}}     
\def\gsim{\mathrel{\rlap{
\lower4pt\hbox{\hskip-3pt$\sim$}}
    \raise1pt\hbox{$>$}}}     
\def\scr#1{\mbox{\scriptsize #1}}
\begin{document}
\title{
Equilibration and 
baryon densities attainable in relativistic heavy-ion collisions 
} 
\author{Yu. B. Ivanov}\thanks{e-mail: yivanov@theor.jinr.ru}
\affiliation{Bogoliubov Laboratory of Theoretical Physics, JINR, Dubna 141980, Russia}
\affiliation{National Research Nuclear University ``MEPhI'', Moscow 115409, Russia}
\affiliation{National Research Centre ``Kurchatov Institute'', 123182 Moscow, Russia} 
\author{A. A. Soldatov}
\affiliation{National Research Nuclear University ``MEPhI'',   
Moscow 115409, Russia}
\begin{abstract}
Kinetic equilibration of the matter and baryon densities 
attained in central region of colliding Au+Au nuclei
in the energy range of $\sqrt{s_{NN}}=$ 3.3--39 GeV 
are examined within the model of the three-fluid dynamics. 
It is found that the kinetic equilibration is faster at higher collision 
energies: the equilibration time (in the c.m. frame of colliding nuclei) 
rises from $\sim$5 fm/c at $\sqrt{s_{NN}}=$ 3.3 GeV to $\sim$1 fm/c at 39 GeV. 
The chemical equilibration, and thus thermalization, takes longer.  
We argue that the presented time evolution of the net-baryon and energy densities in the central region is a necessary prerequisite of proper reproduction of bulk observables in midrapidity. 
We suggest that for informative comparison of predictions of different models 
it is useful to calculate an invariant 4-volume ($V_4$), 
where the proper density the equilibrated  matter exceeds certain value. 
The advantage of this 4-volume is that it does not depend on specific choice of the 
3-volume in different studies and   
takes into account the lifetime of the high-density region, which also matters.  
The 4-volume $V_4=$ 100 fm$^4$/c is chosen 
to compare the baryon densities attainable at different different energies. 
It is found that the highest proper baryon density increases with the collision 
energy rise, from $n_B/n_0\approx$ 4 at 3.3 GeV to $n_B/n_0\approx$ 30 at 39 GeV. 
These highest densities are achieved in the central region of colliding system.
\end{abstract}
\maketitle

\section{Introduction}

The main goal of high-energy heavy-ion research is to explore the properties of strongly 
interacting matter, particularly its phase structure. Initial and final stages of the 
heavy-ion collisions are non-equilibrium. The interest is mainly focused on 
properties of the equilibrated matter, which take place in the intermediate stage of the  
collisions and its evolution is frequently described by hydrodynamical models. 
One of the main questions is what energy and baryon densities can be accessed by means of heavy-ion collisions?
At LHC and top RHIC collision energies systems with a
very small net baryon density but rather high temperature are formed, while
it is expected that the creation of the high 
baryon densities occurs at more moderate collision energies, such as those 
at the  Nuclotron-based Ion Collider fAcility (NICA) 
in Dubna \cite{Kekelidze:2017ghu} and the Facility for Antiproton and Ion Research (FAIR) 
in Darmstadt \cite{Ablyazimov:2017guv} under construction, at the low-energy end of the Relativistic Heavy Ion Collider (RHIC), i.e. the Beam Energy Scan (BES) program at RHIC, 
and at planned J-PARC-HI facility \cite{Sakaguchi:2019xjv}. 
In the present paper we are interested precisely in these moderate-energy heavy-ion collisions.

The question of highest attainable energy and baryon densities in the NICA-FAIR 
energy range was first addressed in Ref. \cite{Randrup07}, where predictions of 
different models where compared. However, the initial equilibration of the matter 
was not analyzed in Ref. \cite{Randrup07}. Besides, since the time of 
Ref. \cite{Randrup07} the models themselves were refined based on numerous 
experimental data from BES RHIC. Therefore, it is reasonable to repeat and extend 
the analysis of Ref. \cite{Randrup07}. Within the 
Ultra-relativistic Molecular Dynamics (UrQMD) model \cite{Bass:1998ca,Bleicher:1999xi}
and the Quark-Gluon String Model (QGSM) \cite{Amelin:1989vp,Amelin:1993xg}
this was done in Refs. \cite{Bravina:2008ra,De:2015hpa} at NICA and FAIR energies. 
The question of highest attainable energy and baryon densities is closely 
related to degree of the baryon stopping in nuclear collisions,  
the discussion of which was recently resumed in Ref. \cite{Mohs:2019iee}.

In the present paper we present the analysis within the 
model of the three-fluid dynamics (3FD) \cite{3FD} in a wider, 
i.e. NICA/FAIR/BES-RHIC,  energy range. Since the time of 
Ref. \cite{Randrup07} the 3FD model \cite{Ivanov:2013wha} was supplemented by 
two equations of state (EoS) involving the deconfinement
 transition \cite{Toneev06}, i.e. a first-order phase transition 
and a smooth crossover one, which turned out to be the most successful in 
reproduction of various observables. Such kind of analysis has been already 
started in Refs. \cite{Ivanov:2017xee,Ivanov:2018rrb}. In the present paper 
we report a more quantitative results for the central region of the 
colliding system.

\section{The 3FD Model}
\label{Model}

The main part of the hydro models 
\cite{Petersen:2008dd,Karpenko:2015xea,Khvorostukhin:2016qxc,Okai:2017ofp,Shen:2017bsr,Akamatsu:2018olk,Du:2018mpf},
that are designed for describing  evolution of the baryon-rich matter,    
takes their initial conditions from third-party kinetic codes.    
Unlike those hybrid hydro models, the 3FD model \cite{3FD}
takes into account finite stopping power of nuclear matter right within the 
3FD evolution. The finite stopping power results in a counterstreaming
regime of leading baryon-rich matter. This nonequilibrium regime 
is modeled by two interpenetrating baryon-rich fluids 
initially associated with constituent nucleons of the projectile
(p) and target (t) nuclei. In addition, newly produced particles
are attributed to a fireball (f) fluid.
Each of these fluids is governed by conventional hydrodynamic equations 
coupled by friction terms in the right-hand sides of the Euler equations. 
These friction terms describe energy--momentum loss of the 
baryon-rich fluids. 
A part of this
loss is transformed into thermal excitation of these fluids, while another part 
gives rise to the particle production into the fireball fluid.
Thus, 
the 3FD approximation is a minimal way to simulate the early-stage nonequilibrium at
high collision energies. 
%
Similar concepts were used in recently developed hybrid models  
\cite{Okai:2017ofp,Shen:2017bsr,Akamatsu:2018olk,Du:2018mpf}. 
Unlike the 3FD, these hybrid models deal with a single equilibrated fluid that however 
does not involve all the matter of colliding nuclei. 
Therefore, this kind of hybrid hydrodynamics contains source terms describing gain of the equilibrated matter in the course of the collision. 
This is similar to the production of the f-fluid in the 3FD.

The counterstreaming of 
the p and t fluids takes place only at the initial stage of
the nuclear collision.   
At later stages the baryon-rich (p and t) fluids have already 
either partially passed though each other or 
partially stopped and unified in the central region. 
The f-fluid also is entrained by the the unified baryon-rich fluid 
but is not that well unified with the latter, thus keeping its 
identity even after the initial unification of the baryon-rich fluids. 
In particular, the friction between
the baryon-rich and net-baryon-free fluids is the
only source of dissipation at the expansion stage. 

The physical input of the present 3FD calculations is described in
Ref.~\cite{Ivanov:2013wha}. 
The 3FD simulations were performed with three different 
equations of state (EoS's): a purely hadronic EoS \cite{gasEOS}  
and two versions of the EoS with the   deconfinement
 transition \cite{Toneev06}, i.e. the first-order phase transition (1PT)  
and crossover one. In the present paper we demonstrate results only with 
the  1PT  and crossover EoS's as the most successful in reproduction of various 
observables in the considered energy range.

\begin{figure}[!h]
\includegraphics[width=6.2cm]{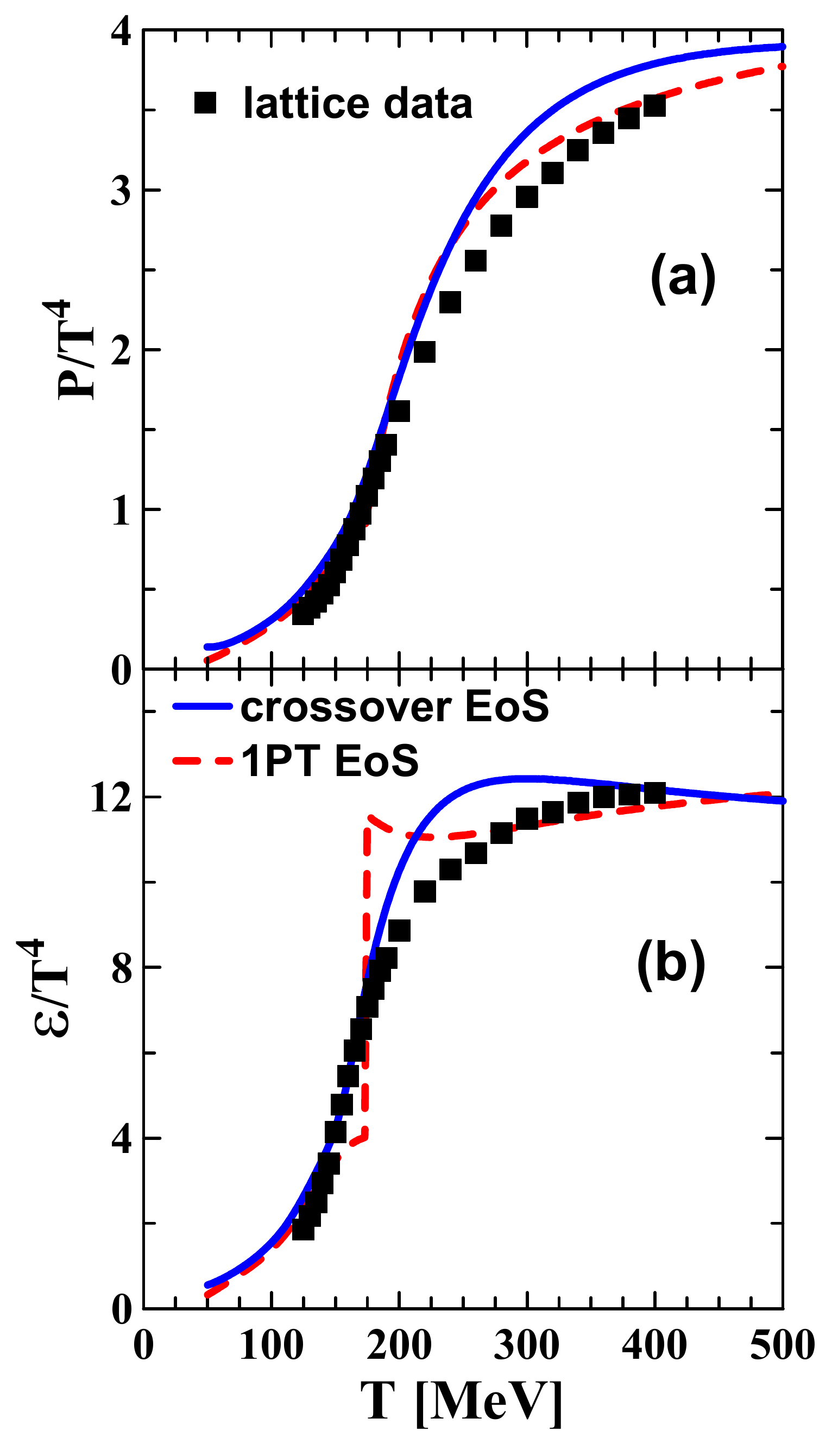}
 \caption{(Color online)
Pressure (a) and energy density (b) as functions of temperature at 
zero baryon chemical potential for the crossover and 1PT EoS's. The
QCD lattice data are from Ref. \cite{Borsanyi:2012cr}.  }
\label{eos-1}
\end{figure}
\begin{figure}[!h]
\includegraphics[width=6.2cm]{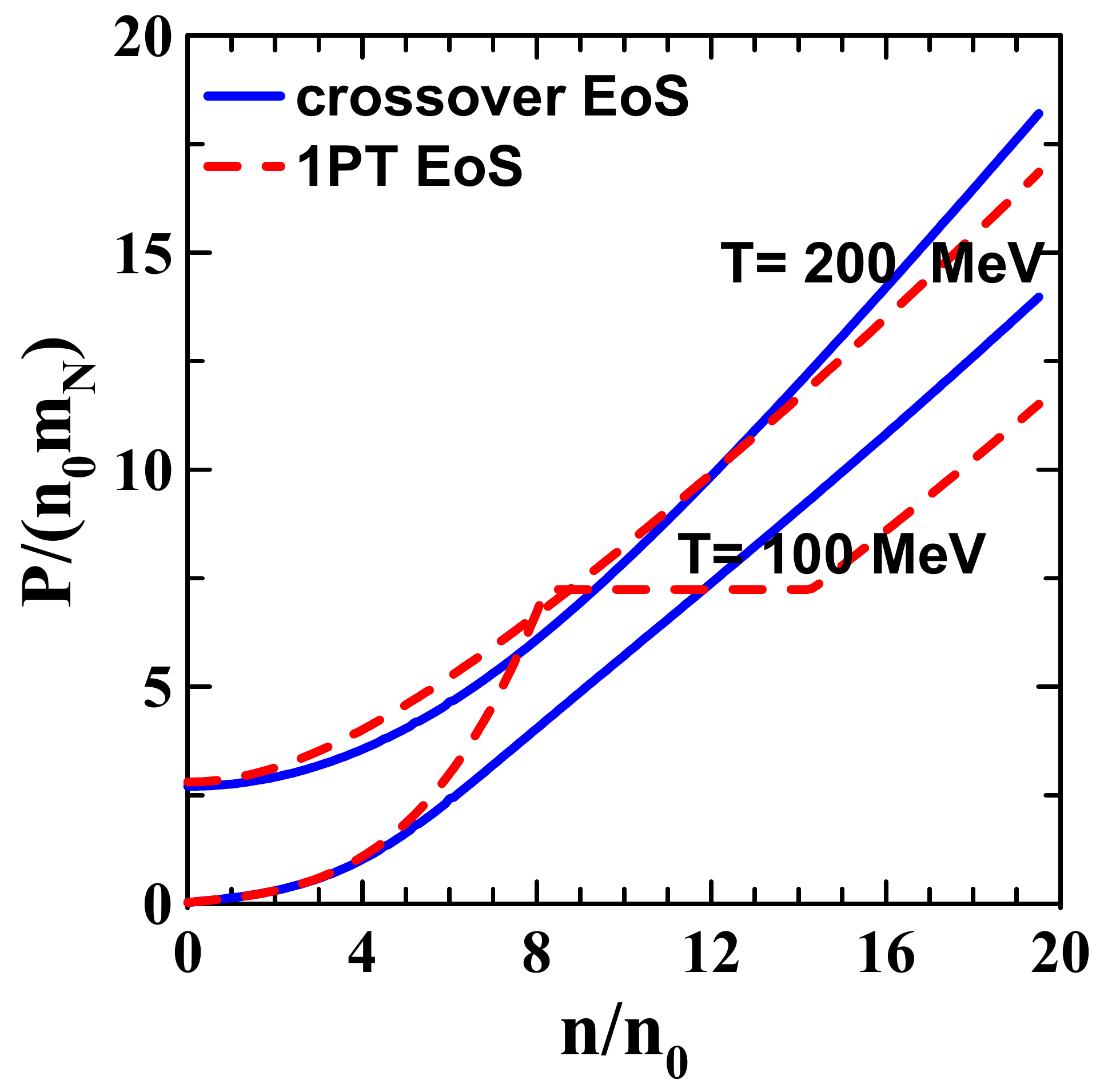}
 \caption{(Color online)
Pressure (scaled by product of the normal nuclear density, $n_0=$ 0.15 1/fm$^3$, 
and the nucleon mass, $m_N$) as function of the net baryon density 
(scaled by $n_0$) for the crossover and 1PT EoS's.   
}
\label{eos-2}
\end{figure}

The used crossover and 1PT EoS's are illustrated in Figs. \ref{eos-1} and \ref{eos-2}. 
Figure \ref{eos-1} demonstrates comparison of these EoS's with the modern lattice data
\cite{Borsanyi:2012cr}. As seen, the reproduction of the lattice data at 
zero baryon chemical potential is not perfect because the considered EoS's were 
fitted to old, still imperfect lattice data \cite{Fodor:2002sd,Csikor:2004ik,Karsch:2000kv}. 
In particular, because of that the critical temperature is $T_c$ = 173 MeV for the 1PT EoS
and $T_c$ = 183 MeV - for the crossover EoS. These $T_c$ values look too high nowadays.  
However, these shortcomings are not severe for the reproduction of bulk observables
in heavy-ion collisions, as it is argued below. 
At finite net baryon densities and moderate temperatures, 
the crossover and 1PT EoS's are very different, as seen from Fig. \ref{eos-2}. 

In spite of the above shortcomings and differences, the crossover and 1PT EoS's 
are equally successful in reproduction of the bulk observables. The crossover EoS 
is only slightly preferable. This a consequence of fitting of the friction
in the quark-gluon phase (QGP) aimed at reproducing the baryon stopping
at high incident energies, as it is described in Ref. \cite{Ivanov:2013wha} in detail. 

\begin{figure}[!h]
\includegraphics[width=6.cm]{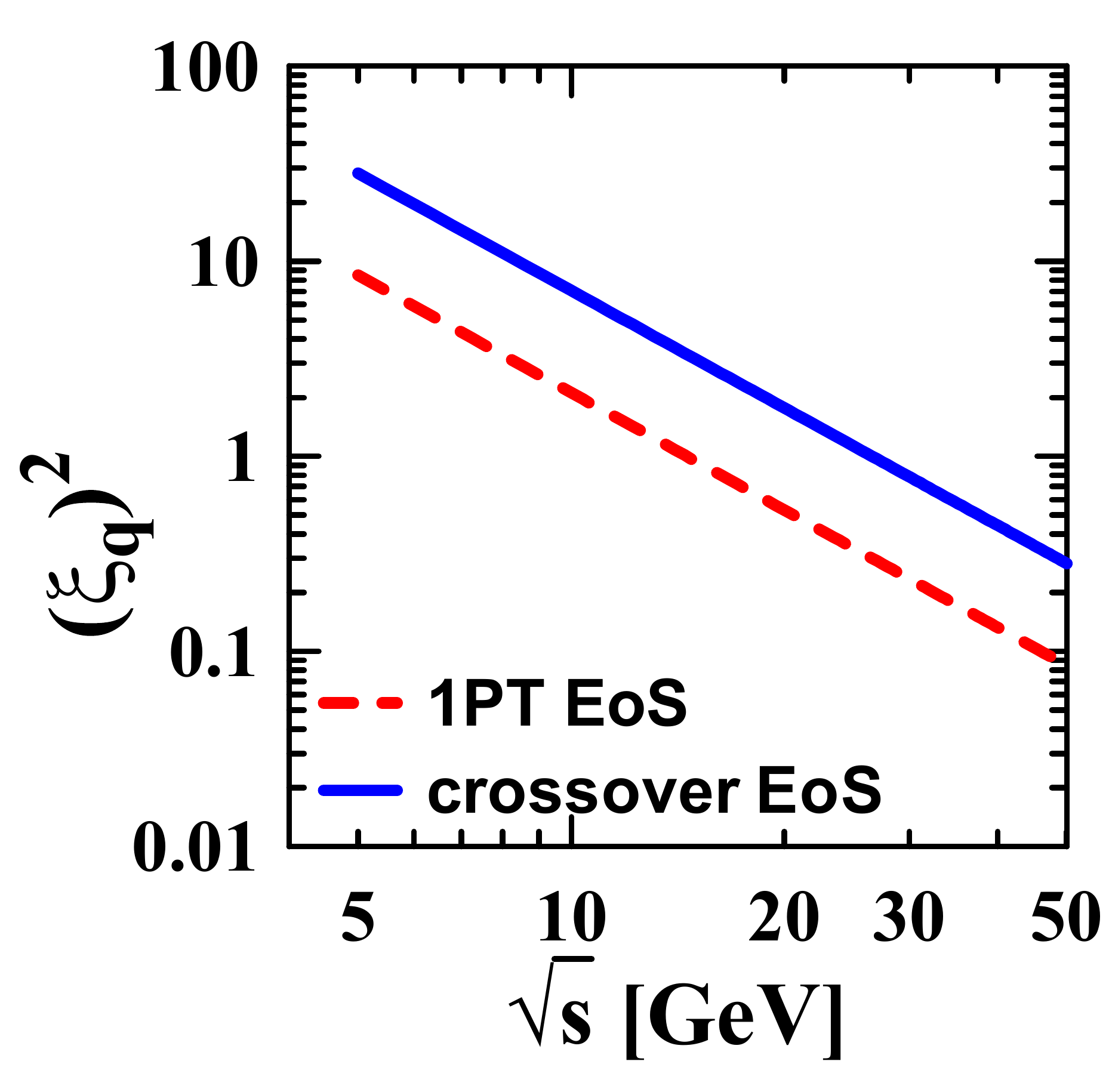}
 \caption{(Color online)
The ratio of the friction in the QGP to  
that in the hadronic phase [$(\xi_q)^2$ in notation of Refs. \cite{3FD,Ivanov:2013wha}, 
see  Eqs. (14) and (16) in Ref. \cite{Ivanov:2013wha}]  
as a function of $\sqrt{s}$, i.e.
the center-of-mass energy of a pair of   
nucleons belonging to the counterstreaming fluids, that locally characterizes 
the  relative velocity of these counterstreaming fluids. 
Fits for the 1PT and crossover EoS's are presented.
}
\label{fig0}
\end{figure}

The friction forces between fluids are key components of the 3FD model, 
which determine the dynamics of a nuclear collision.
In the hadron phase, friction forces were estimated in Ref. \cite{Sat90} 
and have since been used in simulations.
There are no microscopic estimates of
the friction in the QGP so far.
Therefore, the phenomenological friction in the QGP 
was fitted for each EoS separately. 
This fit resulted in the QGP friction, which strongly differs from 
that in the hadronic phase \cite{Sat90} and is quite different for 
the crossover and 1PT EoS's.  
This is illustrated in Fig. \ref{fig0}. 
The weak friction at $\sqrt{s}>$ 20--30 GeV does not imply high transparency.  
At high energies the stopping of the baryon-rich 
counterstreaming fluids results from the friction with the f-fluid that is quite dense at these energies.

Thus, reproduction of the bulk observables depends on a combination 
of the EoS and friction forces rather then only on the EoS. 
Our experience indicates that not any EoS
shortcomings can be cured by fitting the 
friction forces. The tested EoS should be ``not far from the true one''.

\begin{figure*}[!h]
\includegraphics[width=17.cm]{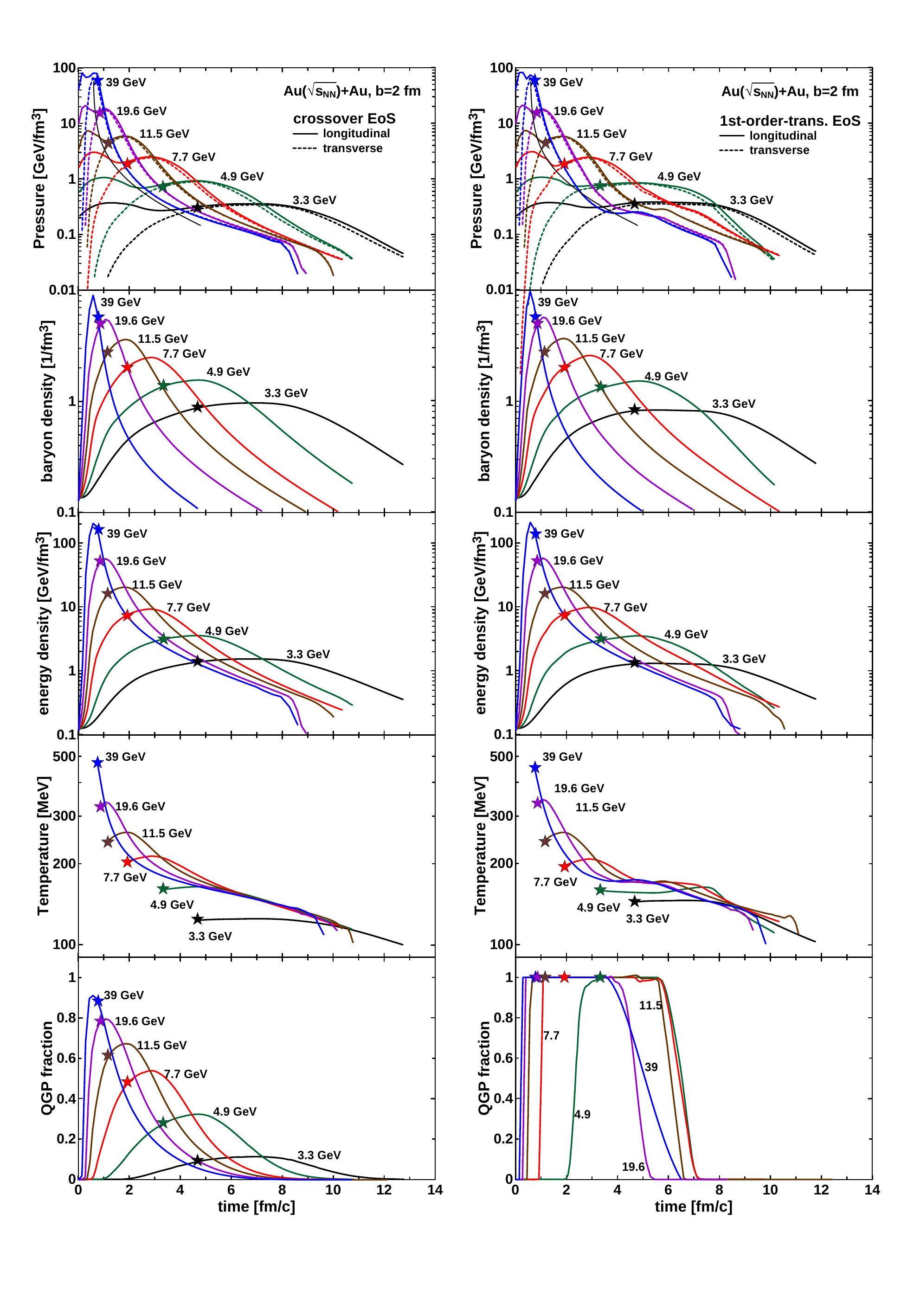}
 \caption{(Color online)
 Time evolution of various quantities 
in the central region of central ($b=$ 2 fm) Au+Au collision at various 
collision energies ($\sqrt{s_{NN}}$). 
From top to bottom: the longitudinal and transverse pressure, net baryon density, 
energy density, temperature, and fraction of the QGP. 
The corresponding quantities are indicated in titles of the  ordinate axes. 
Left column corresponds to the 
crossover EoS, while the right one -- to the 1PT EoS. Star symbols on the curves 
mark the time instant of the equilibration. 
}
\label{fig1a}
\end{figure*}

\section{Matter evolution in central region of colliding nuclei}
\label{central region}

Figure \ref{fig1a} presents evolution of the matter  
in central region of colliding nuclei Au+Au at impact parameter $b=$ 2 fm. 
Similarly to Ref. \cite{Randrup07}, 
the figure displays evolution of various quantities 
in the central box placed around the
origin ${\bf r}=(0,0,0)$ in the frame of equal velocities of
colliding nuclei:  $|x|\leq$ 2 fm,  $|y|\leq$ 2 fm and $|z|\leq$
$\gamma_{cm}$ 2 fm, where $z$ is the direction of the beam
and $\gamma_{cm}$ is the Lorentz
factor associated with the initial nuclear motion in the c.m. frame.  
At high collision energies, the Lorentz contraction can be so high that 
the central box of a fixed (rather than Lorentz contracted) longitudinal 
size may turn out to be half-empty. That would result in incorrect 
determination of densities in such a box. 
The size of the box is chosen 
to be large enough that the amount of matter in it can be
representative to conclude on the medium properties and 
to be small enough to consider the matter in it as a homogeneous
medium. 
The matter in the box still amounts to a minor part
of the total matter of colliding nuclei.  

One of the advantages of this central box is that the matter is at rest in it 
due to symmetry considerations. Therefore, the baryon and energy densities can
be expressed as a direct sum of partial densities of partial densities of 
different fluids 
   \begin{eqnarray}
   \label{nB-prop}
n_B &=& n_{\scr p}u_{\scr p}^{0}+n_{\scr t}u_{\scr t}^{0}+n_{\scr f}u_{\scr f}^{0},   
\\
   \label{e-prop}
\varepsilon &=& \varepsilon_{\scr p}u_{\scr p}^{0}+\varepsilon_{\scr t}u_{\scr t}^{0} 
+\varepsilon_{\scr f}u_{\scr f}^{0}, 
   \end{eqnarray}
where $n_\alpha$ and $\varepsilon_\alpha$ 
are proper (i.e. in the local rest frame) 
baryon and energy densities of different fluids ($\alpha=$ p, t or f), 
respectively, and $u_\alpha^{0}$ stands for the 0-component of the 
hydrodynamic 4-velocity of the $\alpha$-fluid. Notice that $n_{\scr f}\equiv 0$
by construction of the 3FD model, $u_{\scr f}^{0}=1$ and 
$u_{\scr p}^{0}=u_{\scr t}^{0}$ by definition of the central box.

Longitudinal  ($P_{\rm{long}}$) and transverse ($P_{\rm{tr}}$) pressures
\begin{eqnarray}
\label{P_long}
P_{\rm{long}}&=&T_{zz}, \hspace*{5mm}\mbox{(in the beam direction)}, 
\\
P_{\rm{tr}}&=&(T_{xx}+T_{yy})/2
\label{P_tr}
\end{eqnarray}
are defined in terms of the total energy--momentum tensor 
\begin{eqnarray}
\label{T_tot}
T^{\mu\nu} \equiv
T^{\mu\nu}_{\scr p} + T^{\mu\nu}_{\scr t} + T^{\mu\nu}_{\scr f}
\end{eqnarray}
being the sum of conventional hydrodynamical energy--momentum tensors of separate fluids
\begin{eqnarray}
\label{T_alpha}
T_\alpha^{\mu\nu} = (\varepsilon_\alpha + P_\alpha)
u_\alpha^\mu u_\alpha^\nu + g^{\mu\nu} P_\alpha. 
\end{eqnarray}

The initial stage of the nuclear collision is nonequilibrium. 
This is manifested in the fact that the longitudinal  ($P_{\rm{long}}$) and 
transverse ($P_{\rm{tr}}$) pressures are different, see upper panels in 
Fig. \ref{fig1a}.  
Criterion of the equilibration, that we use in the present paper, 
is equality of longitudinal  and transverse pressures
with the accuracy no worse than 10\%. Time instants, when the equilibration 
happens, are marked by star symbols on the curves in Fig. \ref{fig1a}. 
As seen, the nonequilibrium stage lasts from $\sim$ 5 fm/c at $\sqrt{s_{NN}}=$ 3.3 GeV 
to $\sim$ 1 fm/c at collision energy of 39 GeV. This is somewhat shorter that 
reported in the previous study of Ref. \cite{Ivanov:2018rrb}. 
The reason is that equilibration criterion in \cite{Ivanov:2018rrb} was 
the equality of longitudinal  and transverse pressures
with the accuracy {\it better} than 10\%, which differs from ``{\it no worse}'' 
in the present study. The $P_{\rm{long}}$ and $P_{\rm{tr}}$ values become equal 
with the accuracy $\sim$10\% and remain at this level for some time. 
This is why the ``{\it no worse}'' criterion gives shorter termalization times 
than the {\it ``better''} criterion. In fact, the presently used ``{\it no worse}''
criterion is natural for the chosen size of the box. The matter in this box is not 
perfectly homogeneous, which results in a certain difference in 
the $P_{\rm{long}}$ and $P_{\rm{tr}}$ values. Note that  similar $\sim$10\% 
difference appear even at later stages of the expansion.

We consider only so-called kinetic equilibration, characterized the equality of 
longitudinal  and transverse components of the pressure. Because 
of that we does not call it thermalization. The chemical equilibration 
takes longer time, if ever, because the f-fluid retains 
its identity even after the unfication of the baryon-rich fluids. 
This is in agreement with the analysis within the UrQMD  and QGSM models
\cite{Bravina:2008ra,De:2015hpa}, which indicates that the complete thermalization 
(including the chemical one) requires a longer time. 
The kinetic equilibration also take longer in the UrQMD  and QGSM models.

\begin{figure}[!h]
\includegraphics[width=7.cm]{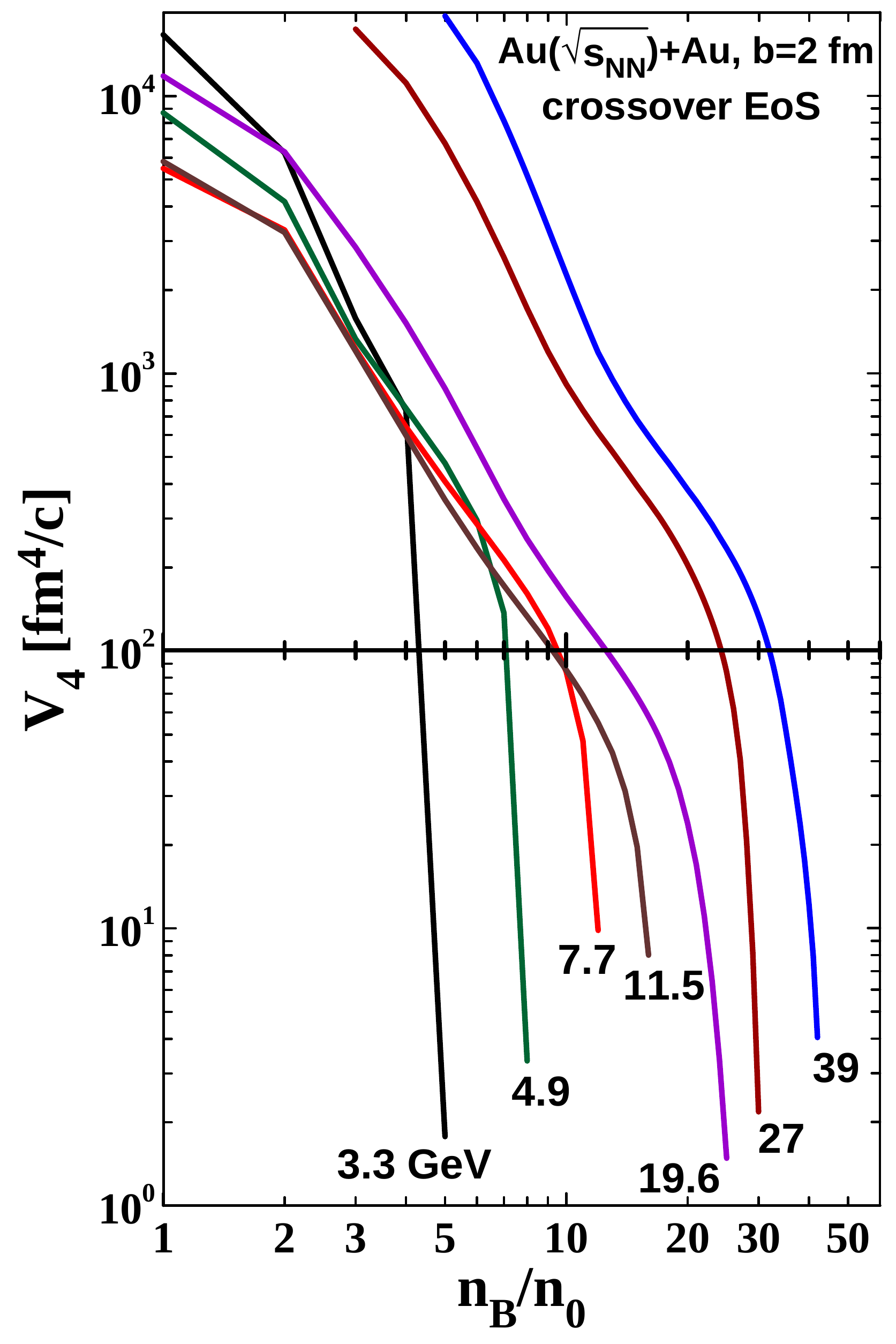}
 \caption{(Color online)
 4-volume, in which the net-baryon density of the equilibrated matter exceeds value $n_B$
(in $n_0$ units, $n_0 =$ 0.15 1/fm$^3$) 
in the central ($b$ = 2 fm) Au+Au collisions at various collision energies
$\sqrt{s_{NN}}$.  Calculations are done with the crossover EoS. 
}
\label{fig1}
\end{figure}
\begin{figure}[!h]
\includegraphics[width=6.5cm]{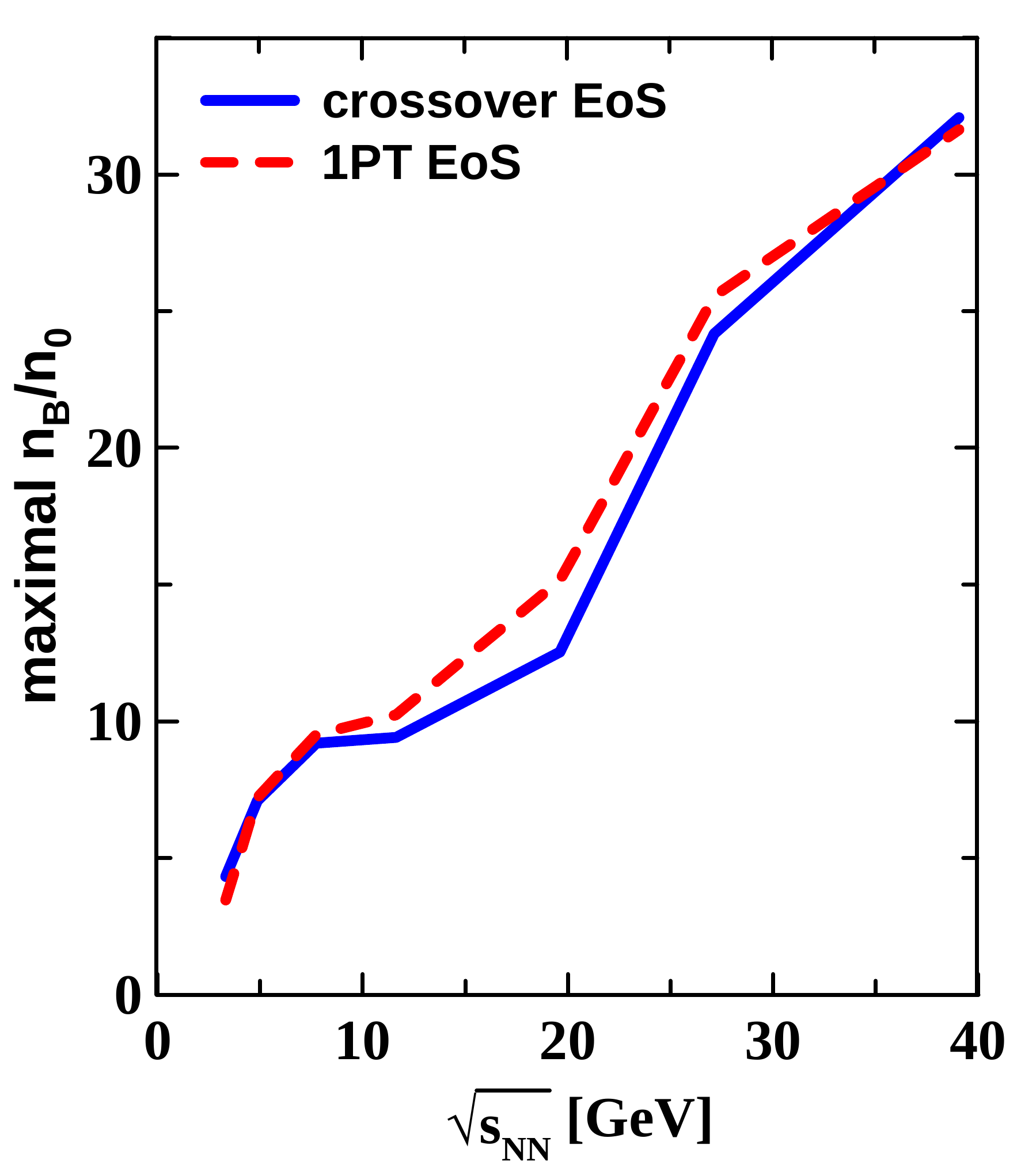}
 \caption{(Color online)
The baryon densities of the equilibrated matter reached in the 4-volume 100 fm$^4$/c 
in the central ($b$ = 2 fm) Au+Au collisions at various collision energies
$\sqrt{s_{NN}}$.
}
\label{fig2}
\end{figure}

The second and third rows of the columns in Fig. \ref{fig1a} present evolution the 
(net) baryon and energy densities in the central box. These densities reach very high 
values, which are again higher than those reported in Ref. \cite{Ivanov:2018rrb}. 
This is again a consequence of the earlier equilibration because of 
above mentioned modification of the corresponding criterion.  However, these high values 
of the densities are not very meaningful because they are achieved in a tiny volume and 
maintained for a tiny     
period of time. It is desirable to evaluate the density in a reasonably large volume and 
which survives for a reasonably long  period of time. Because the 3-volume and the time period
are non-invariant quantities by themselves, following Refs. \cite{Friman:1997sv,3FD}
we define a 4-volume, where the density $n_B$ 
of the {\em equilibrated matter} exceeds certain value $\widetilde{n}_B$
\begin{eqnarray}
\label{V4-all}
V_4 (\widetilde{n}_B) = \int d^4 x \;\Theta \left(n_B(x)-\widetilde{n}_B\right),  
\end{eqnarray}
where $\Theta (x)$ is the step function being equal 1 for $x>0$ and 0 otherwise. 
The $V_4 (\widetilde{n}_B)$ is an invariant measure of the
space--time region, where the $n_B$ value remains high. Note that the 
calculation of $V_4$ is not restricted by the central box described above. 
Technically the $V_4$ computation proceeds as follows. At each time step 
of the computation we sum up 3-volumes of all cells, in which the  
net baryon density exceeds the value of $\widetilde{n}_B$, and multiply the 
resulting 3-volume by the time step. Thus obtained 4-volumes at each time step 
are again summed up.

The $V_4$ values for different collision energies $\sqrt{s_{NN}}$
are presented in Fig. \ref{fig1}. As seen, the $V_4$ values rapidly decrease 
with the rise of $\widetilde{n}_B$. We choose the level of $V_4=$ 100 fm$^4$/c 
to compare the baryon densities attainable at different different energies. 
This 4-volume can be viewed, for example, as 5$\times$5$\times$2fm$^3$$\times$2fm/c, 
i.e. consisting of quite reasonable 3-volume and time period. 
The baryon densities reached in the 4-volume not smaller than 100 fm$^4$/c
are displayed Fig. \ref{fig2}. 
As seen, a very high density is attained at the collision density of 39 GeV, 
this is not surprising because 70\% of the baryon charge gets stopped in 
the central Au+Au collision at this energy, as it is argued 
in Refs. \cite{Ivanov:2017xee,Ivanov:2018rrb}.

The maximal densities achieved in collisions at NICA-FAIR energies are comparable with 
those found within the UrQMD and QGSM models \cite{Bravina:2008ra} in a small central 
cell of 0.5$\times$0.5$\times$0.5 fm$^3$
size while considerably higher that those 
\cite{Bravina:2008ra} in a large central 
cell of 5$\times$5$\times$5 fm$^3$. In contrast to the 3FD results, 
these highest values are achieved at the 
nonequilibrium stage in terms of Ref. \cite{Bravina:2008ra}, because the chemical 
equilibrium is still absent. Lower baryon densities are reported in Ref. Ref. \cite{De:2015hpa}
because the analysis was performed in a larger cell 2$\times$2$\times$2 fm$^3$. 
Thus, results obtained within the same UrQMD model \cite{Bravina:2008ra,De:2015hpa}
cannot be directly compared. Moreover, the observation that the maximal baryon density 
does not rise in the large box of Ref. \cite{Bravina:2008ra} and moderately large box of Ref.
\cite{De:2015hpa} may signal that a box of a fixed longitudinal size contain 
the main part of the colliding system (in longitudinal direction) at the stage of the 
maximal compression. 
Therefore, it is necessary to use the same central box (and better a Lorentz contracted
one) for informative comparison of predictions of different models. 
It would be even better if the analysis is performed in terms of the invariant 4-volume
described above.

For completeness, we also present the evolution of temperature in the fourth 
row of the columns in Fig. \ref{fig1a}.  It is displayed only from the moment of equilibration, 
i.e. when it has a physical sense. One should keep in mind that the temperature
is a EoS-dependent quantity. In the regions, where the QGP dominates 
in both considered scenarios  
(see the bottom row of panels in Fig. \ref{fig1a}), 
the temperatures are also very similar for both EoS's.  
If the phases are different, the temperatures are also different even in spite of 
very similar baryon and energy densities. 
In the crossover case, they gradually decrease almost universally for different energies, 
while for the 1PT EoS their time dependence flattens and the temperatures even slightly 
increase to the end of the QGP, as in the case of the 4.9-GeV energy.

The evolution of the pressure and densities is very similar for the 1PT and 
crossover EoS’s, as seen from Fig. \ref{fig1a}, in spite of difference 
of these EoS’s. 
It takes place because the friction forces in the QGP were
independently fitted for each EoS \cite{Ivanov:2013wha} 
in order to reproduce the baryon stopping, 
as it has been already mentioned in sect. \ref{Model}. 
As a byproduct of this fit, we obtain a very similar evolution of the pressure and densities
for two different EoS's. 
Therefore,  the EoS and the friction forces are only means of achieving  
the presented evolution, which directly results in a proper explanation of  
experimental data on bulk observables. 
Had the QGP friction be fixed by some microscopic calculation, we could 
make a choice between between the the 1PT and crossover EoS’s.

The bottom row of panels in Fig. \ref{fig1a} displays evolution of the 
fraction of the quark-gluon phase (QGP). This QGP-fraction evolution 
demonstrates that the considered 1PT and crossover EoS’s are very different. 
The mixed phase is rapidly passed in the 1PT EoS while the QGP-fraction
becomes close to unity only at the highest considered energies 
$\sqrt{s_{NN}}>$ 20 GeV and for a quite short period of time.

\section{Summary}
\label{Summary}

We estimated the baryon densities of the equilibrated matter
achievable in the central region of colliding Au+Au nuclei 
at NICA/FAIR/BES-RHIC collision energies. 
The analysis was performed within the 3FD  model. 
We  analyzed the kinetically equilibrated matter characterized by equality 
of the longitudinal and transverse components of the pressure.  
It is found that the kinetic equilibration is faster at high collision 
energies: the equilibration time (in the c.m. frame of colliding nuclei) 
decreases from $\sim$5 fm/c at $\sqrt{s_{NN}}=$ 3.3 GeV to $\sim$1 fm/c at 39 GeV. 
The chemical equilibration, and thus thermalization, takes longer 
\cite{Bravina:2008ra}. 
The fact that two different EoS’s supplemented by two (also different) versions  of the friction in the QGP result in a very similar time evolution of the net-baryon and energy densities in the central region indicates that this time evolution is a necessary prerequisite of the proper reproduction of bulk observables in the midrapidity.

We argue that for informative comparison of predictions of different models 
it is useful to calculate a 4-volume, where the proper density 
of the equilibrated matter exceeds certain value. 
The advantage of this 4-volume is that it does not depend on specific choice of the 
3-volume in different studies. Different choices of the 3-volume prevent us even from  
direct comparison or results obtained within the same model 
\cite{Bravina:2008ra,De:2015hpa}. 
In addition, the period of time during which a high density exists also matters. 
The 3-volume and the time period are non-invariant quantities, 
while the 4-volume is an invariant quantity.

This  4-volume is calculated as a function of a threshold density. 
To compare the baryon densities attainable at different different energies,
we choose  the level of $V_4=$ 100 fm$^4$/c. 
This 4-volume can be viewed as consisting of quite reasonable 3-volume and time period. 
It is found that the highest proper baryon density increases with the collision 
energy rise, from $n_B/n_0\approx$ 4 at 3.3 GeV to $n_B/n_0\approx$ 30 at 39 GeV. 
These highest densities are achieved in the central region of colliding system, 
as it is indicated by analysis of the dynamics of these collisions 
\cite{Ivanov:2017xee,Ivanov:2018rrb}.

\begin{acknowledgments}  
This work was carried out using computing resources of the federal collective usage center ``Complex for simulation and data processing for mega-science facilities'' at NRC ``Kurchatov Institute'', http://ckp.nrcki.ru/.
Y.B.I. was supported by the Russian Science
Foundation, Grant No. 17-12-01427, and the Russian Foundation for
Basic Research, Grants No. 18-02-40084 and No. 18-02-40085. 
This work 
was partially supported by the Ministry of Science and Education of the Russian Federation, grant No. 3.3380.2017/4.6, and 
the National Research
Nuclear University MEPhI in the framework of the Russian Academic Excellence Project 
(contract No. 02.a03.21.0005, 27.08.2013).  
%
\end{acknowledgments}


\begin{thebibliography}{999}
%
\bibitem{Kekelidze:2017ghu} 
  V.~D.~Kekelidze, V.~A.~Matveev, I.~N.~Meshkov, A.~S.~Sorin and G.~V.~Trubnikov,
  Phys.\ Part.\ Nucl.\  {\bf 48}, no. 5, 727 (2017).
%
\bibitem{Ablyazimov:2017guv} 
  T.~Ablyazimov {\it et al.} [CBM Collaboration],
  Eur.\ Phys.\ J.\ A {\bf 53}, no. 3, 60 (2017)
  [arXiv:1607.01487 [nucl-ex]].
%
\bibitem{Sakaguchi:2019xjv} 
  T.~Sakaguchi [J-PARC-HI Collaboration],
  PoS CORFU {\bf 2018}, 189 (2019)
  [arXiv:1904.12821 [nucl-ex]].
%
\bibitem{Randrup07}
  I.~C.~Arsene {\it et al.},
  Phys.\ Rev.\ C {\bf 75}, 034902 (2007)
  [nucl-th/0609042].
%
\bibitem{Bass:1998ca} 
  S.~A.~Bass {\it et al.},
  Prog.\ Part.\ Nucl.\ Phys.\  {\bf 41}, 255 (1998)
  [nucl-th/9803035].
%
\bibitem{Bleicher:1999xi} 
  M.~Bleicher {\it et al.},
  J.\ Phys.\ G {\bf 25}, 1859 (1999)
  [hep-ph/9909407].
%
\bibitem{Amelin:1989vp} 
  N.~S.~Amelin and L.~V.~Bravina,
  Sov.\ J.\ Nucl.\ Phys.\  {\bf 51}, 133 (1990)
  [Yad.\ Fiz.\  {\bf 51}, 211 (1990)];
%
\bibitem{Amelin:1993xg} 
  N.~S.~Amelin, L.~V.~Bravina, L.~P.~Csernai, V.~D.~Toneev, K.~K.~Gudima and S.~Y.~Sivoklokov,
  Phys.\ Rev.\ C {\bf 47}, 2299 (1993).
%
\bibitem{Bravina:2008ra} 
  L.~V.~Bravina {\it et al.},
  Phys.\ Rev.\ C {\bf 78}, 014907 (2008)
  [arXiv:0804.1484 [hep-ph]].
%
\bibitem{De:2015hpa} 
  S.~De, S.~De and S.~Chattopadhyay,
  Phys.\ Rev.\ C {\bf 94}, no. 5, 054901 (2016)
  [arXiv:1510.01456 [nucl-th]].
%
%
\bibitem{Mohs:2019iee} 
  J.~Mohs, S.~Ryu and H.~Elfner,
  arXiv:1909.05586 [nucl-th].
%
%
\bibitem{3FD}
 Yu. B. Ivanov, V. N. Russkikh, and V.D. Toneev,
 Phys. Rev. C {\bf 73}, 044904 (2006) 
[nucl-th/0503088].
%
\bibitem{Ivanov:2013wha} 
  Yu.~B.~Ivanov,
  Phys. Rev. C {\bf 87}, 064904 (2013) [arXiv:1302.5766 [nucl-th]]. 
%
\bibitem{Toneev06}
A. S. Khvorostukhin,  
V. V. Skokov, K. Redlich, and V. D. Toneev,
Eur. Phys. J. {\bf C48}, 531 (2006)  [nucl-th/0605069].
%
\bibitem{Ivanov:2017xee} 
  Y.~B.~Ivanov and A.~A.~Soldatov,
  Phys.\ Rev.\ C {\bf 97}, no. 2, 021901 (2018)
  [arXiv:1711.03069 [nucl-th]].
%
\bibitem{Ivanov:2018rrb} 
  Y.~B.~Ivanov and A.~A.~Soldatov,
  Phys.\ Rev.\ C {\bf 98}, no. 1, 014906 (2018)
  [arXiv:1803.11474 [nucl-th]].
%
\bibitem{Borsanyi:2012cr} 
  S.~Borsanyi, G.~Endrodi, Z.~Fodor, S.~D.~Katz, S.~Krieg, C.~Ratti and K.~K.~Szabo,
  JHEP {\bf 1208}, 053 (2012)
  [arXiv:1204.6710 [hep-lat]].
%
\bibitem{Fodor:2002sd} 
  Z.~Fodor,
  Nucl.\ Phys.\ A {\bf 715}, 319 (2003)
  [hep-lat/0209101].
%
\bibitem{Csikor:2004ik} 
  F.~Csikor, G.~I.~Egri, Z.~Fodor, S.~D.~Katz, K.~K.~Szabo and A.~I.~Toth,
  JHEP {\bf 0405}, 046 (2004)
  [hep-lat/0401016].
%
\bibitem{Karsch:2000kv} 
  F.~Karsch, E.~Laermann and A.~Peikert,
  Nucl.\ Phys.\ B {\bf 605}, 579 (2001)
  [hep-lat/0012023].
%
%
%
\bibitem{Sat90} L. M.~Satarov, Sov. J. Nucl. Phys. {\bf 52}, 264 (1990).
%
%
\bibitem{Petersen:2008dd} 
  H.~Petersen, J.~Steinheimer, G.~Burau, M.~Bleicher and H.~Stocker,
  Phys.\ Rev.\ C {\bf 78}, 044901 (2008)
  [arXiv:0806.1695 [nucl-th]].
%
\bibitem{Karpenko:2015xea} 
  I.~A.~Karpenko, P.~Huovinen, H.~Petersen and M.~Bleicher,
  Phys.\ Rev.\ C {\bf 91}, no. 6, 064901 (2015)
  [arXiv:1502.01978 [nucl-th]].
%
\bibitem{Khvorostukhin:2016qxc} 
  A.~S.~Khvorostukhin and V.~D.~Toneev,
  Phys.\ Part.\ Nucl.\ Lett.\  {\bf 14}, no. 1, 9 (2017)
  [arXiv:1606.00987 [nucl-th]];
%
  Phys.\ Atom.\ Nucl.\  {\bf 80}, no. 2, 285 (2017)
  [Yad.\ Fiz.\  {\bf 80}, no. 2, 161 (2017)];
%
  A.~S.~Khvorostukhin, E.~E.~Kolomeitsev and V.~D.~Toneev,
  arXiv:1810.10303 [nucl-th].
%
\bibitem{Okai:2017ofp} 
  M.~Okai, K.~Kawaguchi, Y.~Tachibana and T.~Hirano,
  Phys.\ Rev.\ C {\bf 95}, no. 5, 054914 (2017)
  [arXiv:1702.07541 [nucl-th]].
%
\bibitem{Shen:2017bsr} 
  C.~Shen and B.~Schenke,
  Phys.\ Rev.\ C {\bf 97}, no. 2, 024907 (2018)
  [arXiv:1710.00881 [nucl-th]];
  G.~S.~Denicol, C.~Gale, S.~Jeon, A.~Monnai, B.~Schenke and C.~Shen,
  Phys.\ Rev.\ C {\bf 98}, no. 3, 034916 (2018)
  [arXiv:1804.10557 [nucl-th]].
%
\bibitem{Akamatsu:2018olk} 
  Y.~Akamatsu {\it et al.},
  Phys.\ Rev.\ C {\bf 98}, no. 2, 024909 (2018)
  [arXiv:1805.09024 [nucl-th]].
%
\bibitem{Du:2018mpf} 
  L.~Du, U.~Heinz and G.~Vujanovic,
  Nucl.\ Phys.\ A {\bf 982}, 407 (2019)
  [arXiv:1807.04721 [nucl-th]].  
%
%
%
\bibitem{gasEOS}
V. M. Galitsky and I. N. Mishustin, Sov. J. Nucl. Phys. {\bf 29}, 181 (1979).
%
\bibitem{Friman:1997sv} 
  B.~Friman, W.~Norenberg and V.~D.~Toneev,
  Eur.\ Phys.\ J.\ A {\bf 3}, 165 (1998)
  [nucl-th/9711065].
%

\end{thebibliography}
\end{document}